\documentclass{jpsj3}
\usepackage{txfonts}

\title{Microscopic Derivation of Ginzburg-Landau Equations 
 for Coexistent States of  Superconductivity and Magnetism}

\author{\name{Kazuhiro \surname{KUBOKI}}\thanks{E-mail address: kuboki@kobe-u.ac.jp} 
and \name{Keiji \surname{YANO}} 
}
\inst{\address{Department of Physics, Kobe University, 
Kobe 657-8501} 
}

\abst{Ginzburg-Landau (GL) equations for the coexistent states of  superconductivity 
and magnetism are derived microscopically from the extended 
Hubbard model with on-site repulsive and nearest-neighbor attractive interactions.
 In the derived GL free energy a cubic term that couples the  
spin-singlet and spin-triplet components of superconducting order parameters (SCOP) 
with magnetization exists. 
This term gives rise to a spin-triplet SCOP  near the interface 
between a spin-singlet superconductor and a ferromagnet,    
consistent with previous theoretical studies based on the Bogoliubov de Gennes 
method and the quasiclassical Green's function theory.  
In coexistent states of singlet superconductivity and 
antiferromagnetism it leads to the occurrence of $\pi$-triplet SCOPs. 
 }

\kword{GL theory, unconventional superconductivity, coexistence, proximity effect}

\begin{document}
\maketitle

\section{Introduction}
The coexistence and competition of superconductivity and magnetism have been 
important issues in various strongly correlated electron systems, 
{\it e.g.}, high-$T_C$ cuprate superconductors.\cite{Kitaoka}
This is because these two ordered states originate from the same interaction; 
thus, understanding their relation 
may give insight into the mechanism of superconductivity. 

Heterostructures composed of superconductors 
and magnetic materials may be useful systems for studying these problems. 
The properties of the states near an interface strongly depend on the 
materials used, especially the symmetry of superconducting (SC)  
states and the underlying electronic states,\cite{Buz,Berg,Demler,KK1,Kashiwaya,
Zhu,Halterman,Berg2,Esch,Braude,KK2,Krawiec,Brydon,Brydon2,Cuoco}
namely,  the shape of Fermi surfaces and the type of interactions. 
The interface states of heterostructures may have quite different characters 
from those in the bulk. 
Not only the coexistence of the original order parameters (OPs),  
but also new ordered states may arise depending 
on the constituent material.
For example, spin-triplet SCOPs can occur near the interface between 
a spin-singlet superconductor and a ferromagnet. 
This was theoretically 
found using the Bogoliubov de Gennes (BdG) method\cite{KK1} and 
the quasiclassical Green's functions theory. \cite{Berg2,Esch}
Therefore, by combining various types of superconductors with ferromagnets or 
antiferromagnets, we may know the conditions under which 
a particular SC state can be realized.  

In this paper, we derive GL equations and the GL free energy 
microscopically from a tight-binding model on a square lattice with on-site 
repulsive and nearest-neighbor attractive interactions, 
{\it i.e.}, the extended Hubbard model. 
Although this model is a minimal one for treating magnetism and 
unconventional superconductivity,  
it exhibits $s$-, $d$-, and chiral $(p_x \pm ip_y)$-wave
superconductivity,\cite{KK3,Mic} and ferro- and 
anti-ferromagnetism\cite{Hirsch}  for different choices of the  
parameters, especially the electron density 
(in other words, the shape of the Fermi surface).
For this reason this model may be used to examine the material 
dependence of the interface states of heterostructures composed of 
superconductors and magnetic materials. 
The method of deriving GL equations is based on that 
by Gor'kov\cite{Gorkov} with an extension to include magnetic OPs.  
The resulting GL equations are coupled equations for all types of OPs 
including magnetization.  
(The GL equations for superconductors with $s$- and $d$-wave 
SCOPs have already been obtained from a similar model.\cite{Ren,Feder})

Although the GL theory is reliable only qualitatively except near $T_C$, 
it can give a simple and clear description  of the 
coexistence and competition of multiple OPs.
Thus, it is complementary to more sophisticated methods such as 
the BdG and quasiclassical Green's function theory. 

This paper is organized as follows. 
In $\S$2, we present the model and treat it by 
 mean-field approximation (MFA). 
In $\S$3,  GL equations and the GL free energy are derived 
for the coexistent states of superconductivity and ferromagnetism. 
The case of antiferromagnetism and superconductivity is examined in  
$\S$4. Section 5 is devoted to summary and discussion.

\section{Model and Mean-Field Approximation}

We consider the extended Hubbard model on a square lattice, {\it i.e.},  
a tight-binding model that has on-site repulsive and nearest-neighbor 
attractive interactions. (We use the units $\hbar = k_B =1$, and the lattice 
constant is taken to be unity.) 
By treating the  latter interaction  using MFA, 
$s$-, $d$-, and chiral $(p_x \pm ip_y)$-wave SC states 
can be realized depending on the electron density.  
Namely, the symmetry of the SC state may be  
determined by the shape of the Fermi surface. \cite{KK3,Mic}
Similarly, the magnetic order, either ferromagnetic and antiferromagnetic (AF), 
can be obtained by treating the repulsive interaction by MFA.\cite{Hirsch}
 
The Hamiltonian of our model is given by 
\begin{equation}\begin{array}{rl}
H & = \displaystyle -t\sum_{j\sigma} \sum_{\delta={\hat x},{\hat y}}
\big(c^\dagger_{j+\delta,\sigma}c_{j.\sigma}e^{i\phi_{j+\delta,j}} 
+ h.c. \big)
- \mu_0\sum_{j\sigma}c^\dagger_{j\sigma}c_{j\sigma} \\
& \displaystyle +U\sum_j n_{j\uparrow}n_{j\downarrow} 
-V\sum_j\sum_{\delta={\hat x},{\hat y}}
\big(n_{j\uparrow}n_{j+\delta\downarrow} 
+ n_{j\downarrow}n_{j+\delta\uparrow}\big),  
\end{array}\end{equation}
where $t$, $\mu_0$, $U$, and $V$ are the transfer integral, chemical potential, 
and the 
on-site repulsive and nearest-neighbor attractive interactions, respectively; 
$\sigma = \uparrow, \downarrow$ is the spin index.
The magnetic field is taken into account using the Peierls phase 
$\phi_{j,\ell} \equiv \frac{\pi}{\phi_0} \int_j^{\ell}{\bf A}\cdot d{\bf l}$, 
with ${\bf A}$ and $\phi_0 = \frac{hc}{2e}$ being the vector potential 
and  flux quantum, respectively. 
We treat this Hamiltonian using the standard  procedure of MFA:
\begin{equation}\begin{array}{rl}
n_{j\uparrow}n_{j\downarrow} \to & \displaystyle 
\langle n_{j\uparrow}\rangle n_{j\downarrow} 
+\langle n_{j\downarrow}\rangle n_{j\uparrow} 
- \langle n_{j\uparrow}\rangle \langle n_{j\uparrow}\rangle, \\
n_{j\uparrow}n_{\ell\downarrow} \to & \displaystyle 
\Delta_{j,\ell} c^\dagger_{l\downarrow}c^\dagger_{j\uparrow} 
+ \Delta_{j,\ell}^* c_{j\uparrow}c_{\ell\downarrow} 
-|\Delta_{j,\ell}|^2.     
\end{array}\end{equation} 
The SCOPs and magnetization ({\it i.e.}, the OP for magnetism)  
are defined as 
\begin{equation}
\displaystyle \Delta_{j,\ell} = \langle c_{j\uparrow}c_{\ell\downarrow}\rangle, \ \ 
m_j = \frac{1}{2} \langle n_{j\uparrow}-n_{j\downarrow}\rangle.
\end{equation} 
Then the mean-field Hamiltonian is written as 
\begin{equation}\begin{array}{rl}
H_{MF} & = \displaystyle -t\sum_{j\sigma} \sum_{\delta=\pm{\hat x},\pm{\hat y}}
c^\dagger_{j+\delta,\sigma}c_{j,\sigma} e^{i\phi_{j+\delta,j}} \\
& \displaystyle +\sum_j \big[\big(-\mu+Um_j\big)
c^\dagger_{j\downarrow}c_{j\downarrow} 
- \big(\mu+Um_j\big)c^\dagger_{j\uparrow}c_{j\uparrow}\big] \\
& \displaystyle - V\sum_j\sum_{\delta=\pm{\hat x},\pm{\hat y}}
\big[\Delta_{j,j+\delta}c^\dagger_{j+\delta\downarrow}c^\dagger_{j\uparrow}
+ h.c. \big] + E_0, 
\end{array}\end{equation} 
where 
\begin{equation}
E_0 = \displaystyle U\sum_j\big[m_j^2-\frac{1}{4}(n^{(0)}_j)^2\big]
+ V\sum_j\sum_{\delta=\pm{\hat x},\pm{\hat y}}|\Delta_{j.j+\delta}|^2, 
\end{equation}
with $n^{(0)}_j = \langle n_{j\uparrow}+n_{j\downarrow} \rangle$  being 
the electron density at the site $j$. 
Here, $\mu = \mu_0 -Un_0/2$ is the renormalized chemical potential
with $n_0$ being the average electron density of the system. 

In order to derive the GL equations for OPs, we introduce the following 
thermal Green's functions:  
\begin{equation}
G_\sigma(j,\ell,\tau) =  \displaystyle 
-\langle T_\tau c_{j\sigma}(\tau)c_{\ell\sigma}^\dagger\rangle,  \ \ 
F^\dagger_{\sigma\sigma'}(j,\ell,\tau) = 
-\langle T_\tau c_{j\sigma}^\dagger(\tau)c_{\ell\sigma'}^\dagger\rangle.  
\end{equation} 
The equations of motion  for $G_\sigma$ and $F^\dagger_{\sigma\sigma'}$ 
({\it i.e.}, the Gor'kov equations) are obtained by 
taking their $\tau$ derivatives and carrying out Fourier transformation to the Matsubara 
frequency $i\varepsilon_n$ ($=(2n+1)i\pi T$; $T$ being the temperature).  
%
%
%
%
%
%
These equations can be transformed to the following coupled 
equations for $G_\sigma$ and $F^\dagger_{\sigma\sigma'}$:  
\begin{equation}\begin{array}{rl}
 G_\uparrow(j,\ell,i\varepsilon_n) = & \displaystyle {\tilde G}_0(j,\ell,i\varepsilon_n)  
 + V\sum_{k,\delta} {\tilde G}_0(j,k,i\varepsilon_n) \Delta_{k,k+\delta} 
F^\dagger_{\downarrow\uparrow}(k+\delta,\ell,i\varepsilon_n) 
\\ & \displaystyle 
-U\sum_k {\tilde G}_0(j,k,i\varepsilon_n) m_k 
G_\uparrow(k,\ell,i\varepsilon_n), \\ 

G_\downarrow(j,\ell,i\varepsilon_n) = 
& \displaystyle {\tilde G}_0(j,\ell,i\varepsilon_n) 
- V\sum_{k,\delta} {\tilde G}_0(j,k,i\varepsilon_n) \Delta_{k+\delta,k} 
F^\dagger_{\uparrow\downarrow}(k+\delta,\ell,i\varepsilon_n) \\
& \displaystyle +U\sum_k {\tilde G}_0(j,k,i\varepsilon_n) m_k 
G_\downarrow(k,\ell,i\varepsilon_n), \\ 

F^\dagger_{\downarrow\uparrow}(j,\ell,i\varepsilon_n) = & \displaystyle
-V\sum_{k,\delta} {\tilde G}_0(k,j,-i\varepsilon_n) \Delta^*_{k+\delta,k} 
G_\uparrow(k+\delta,\ell,i\varepsilon_n) \\
& \displaystyle +U\sum_k {\tilde G}_0(k,j,-i\varepsilon_n) m_k 
F^\dagger_{\downarrow\uparrow}(k,\ell,i\varepsilon_n),  \\ 

F^\dagger_{\uparrow\downarrow}(j,\ell,i\varepsilon_n) = & \displaystyle
V\sum_{k,\delta} {\tilde G}_0(k,j,-i\varepsilon_n) \Delta^*_{k.k+\delta} 
G_\downarrow(k+\delta,\ell,i\varepsilon_n) \\
& \displaystyle -U\sum_k {\tilde G}_0(k,j,-i\varepsilon_n) m_k 
F^\dagger_{\uparrow\downarrow}(k,\ell,i\varepsilon_n),  \\ 
\end{array}\end{equation} 
where the summation on $\delta$ ($k$) is over 
 $\pm {\hat x}$ and $\pm {\hat y}$ (all sites). 
Here,  ${\tilde G}_0(j,\ell,i\omega_n)$  is  Green's function for
 the system without $\Delta$ and $m$  but with ${\bf A}$ satisfying 
\begin{eqnarray}
\displaystyle (i\varepsilon_n+\mu) {\tilde G}_0(j,\ell,i\varepsilon_n) 
+ t\sum_\delta{\tilde G}_0(j+\delta,\ell,i\varepsilon_n)  
e^{i\phi_{j,j+\delta}} = \delta_{j,\ell}.
\end{eqnarray}
${\tilde G}_0$ is related to  Green's function for the system without ${\bf A}$, 
$G_0$, as 
$
 {\tilde G}_0(j,\ell,i\varepsilon_n) 
=   G_0(j,\ell,i\varepsilon_n) e^{i\phi_{j,\ell}}$. 
$G_0(j,\ell,i\varepsilon_n)$ is the Fourier transform of 
$
G_0({\bf p},i\varepsilon_n) = 1/(i\varepsilon_n-\xi_p)$ with 
$\displaystyle \xi_p = -2t(\cos p_x+\cos p_y)-\mu$. 
%

Spin-singlet and spin-triplet SCOPs on the bond $(j,j+\eta)$ are expressed 
in terms of  Green's functions $F_{\uparrow\downarrow}^\dagger$ 
and $F_{\downarrow\uparrow}^\dagger$, 
\begin{equation}\begin{array}{rl}
\displaystyle (\Delta_\eta^{(S)}(j))^* \equiv   & \displaystyle \frac{1}{2} \langle 
c_{j\uparrow}c_{j+\eta\downarrow} - c_{j\downarrow}c_{j+\eta\uparrow}\rangle^*  
=   \frac{1}{2} \big(\Delta_{j,j+\eta}+\Delta_{j+\eta,j}\big)^* \\
= & \displaystyle \frac{1}{2} T\sum_{\varepsilon_n}
\Big[F^\dagger_{\uparrow\downarrow}(j+\eta,j,i\varepsilon_n) 
- (F^\dagger_{\downarrow\uparrow}(j+\eta,j,i\varepsilon_n)\Big], \\
\displaystyle  (\Delta_\eta^{(T)}(j))^* \equiv & \displaystyle\frac{1}{2} \langle 
c_{j\uparrow}c_{j+\eta\downarrow} + c_{j\downarrow}c_{j+\eta\uparrow}\rangle^*
=   \frac{1}{2} \big(\Delta_{j,j+\eta}-\Delta_{j+\eta,j} \big)^* \\
= &  \displaystyle  -\frac{1}{2} T\sum_{\varepsilon_n}
\Big[F^\dagger_{\uparrow\downarrow}(j+\eta,j,i\varepsilon_n) 
+ F^\dagger_{\downarrow\uparrow}(j+\eta,j,i\varepsilon_n)\Big], \\ 
\end{array}\end{equation}
and the magnetization is similarly given using  $G_{\uparrow}$ 
and $G_{\downarrow}$ as 
\begin{equation}\begin{array}{rl}
 m_j \equiv & \displaystyle \frac{1}{2} \langle c^\dagger_{j\uparrow}c_{j\uparrow} 
- c^\dagger_{j\downarrow}c_{j\downarrow} \rangle \\
= & \displaystyle \frac{1}{2} T\sum_{\varepsilon_n}
\big[G_\uparrow(j,j,i\varepsilon_n)-G_\downarrow(j,j,i\varepsilon_n) \big]. 
\end{array}\end{equation}
We substitute eq. (7) into eqs. (9) and (10) iteratively 
and keep the terms up to the third order in OPs 
to get the following GL equations:   
\begin{equation}\begin{array}{rl}
\displaystyle (\Delta^{(S)}_\eta(j))^* =  & \displaystyle 
 \sum_{k,\delta}L^{(1)}(j,k,\eta,\delta)(\Delta_\delta^{(S)}(k))^* 
+ \sum_{k.k',\delta}L^{(2)}(j,k,k',\eta,\delta)(\Delta_\delta^{(T)}(k))^* m_{k'} 
\\ 
+ & \displaystyle \sum_{k,k',k''}\sum_{\delta,\delta',\delta''}
 L^{(3)}(j,k,k',k'',\eta,\delta,\delta',\delta'')
 \Big[(\Delta_\delta^{(S)}(k))^*(\Delta_{\delta'}^{(S)}(k'))
 (\Delta_{\delta''}^{(S)}(k''))^*
  \\ & \displaystyle 
 - (\Delta_\delta^{(S)}(k))^*(\Delta_{\delta'}^{(T)}(k'))
 (\Delta_{\delta''}^{(T)}(k''))^* 
 -  (\Delta_\delta^{(T)}(k))^*(\Delta_{\delta'}^{(T)}(k'))
 (\Delta_{\delta''}^{(S)}(k''))^*
 \\ & \displaystyle 
+  (\Delta_\delta^{(T)}(k))^*(\Delta_{\delta'}^{(S)}(k'))
 (\Delta_{\delta''}^{(T)}(k''))^*\Big] 
 \\ 
+ & \displaystyle \sum_{k,k',k'',\delta}L^{(4)}(j,k,k',k'',\eta,\delta) 
(\Delta_\delta^{(S)}(k))^*m_{k'}m_{k''}, 
 \end{array}\end{equation}
 \begin{equation}\begin{array}{rl}
\displaystyle (\Delta^{(T)}_\eta(j))^*   = 
- & \displaystyle \sum_{k,\delta}L^{(1)}(j,k,\eta,\delta)(\Delta_\delta^{(T)}(k))^* 
-\sum_{k.k',\delta}L^{(2)}(j,k,k',\eta,\delta)(\Delta_\delta^{(S)}(k))^* m_{k'} 
\\ 
+& \displaystyle \sum_{k,k',k''}\sum_{\delta,\delta',\delta''}
 L^{(3)}(j,k,k',k'',\eta,\delta,\delta',\delta'')
 \Big[(\Delta_\delta^{(T)}(k))^*(\Delta_{\delta'}^{(T)}(k'))
 (\Delta_{\delta''}^{(T)}(k''))^*
  \\ & \displaystyle 
 - (\Delta_\delta^{(T)}(k))^*(\Delta_{\delta'}^{(S)}(k'))
 (\Delta_{\delta''}^{(S)}(k''))^* 
 -  (\Delta_\delta^{(S)}(k))^*(\Delta_{\delta'}^{(S)}(k'))
 (\Delta_{\delta''}^{(T)}(k''))^*
 \\ & \displaystyle 
+  (\Delta_\delta^{(S)}(k))^*(\Delta_{\delta'}^{(T)}(k'))
 (\Delta_{\delta''}^{(S)}(k''))^*\Big] 
 \\ 
- & \displaystyle \sum_{k,k',k'',\delta}L^{(4)}(j,k,k',k'',\eta,\delta) 
(\Delta_\delta^{(T)}(k))^*m_{k'}m_{k''}, 
 \end{array}\end{equation}
 \begin{equation}\begin{array}{rl}
m_j = & \displaystyle \sum_k L^{(5)}(j,k) m_k 
+ \sum_{k,k',\delta,\delta'} L^{(6)}(j,k,k',\delta,\delta')  
\Big[\Delta_\delta^{(S)}(k)(\Delta_{\delta'}^{(T)}(k'))^*
- \Delta_\delta^{(T)}(k)(\Delta_{\delta'}^{(S)}(k'))^*\Big] 
 \\ & \displaystyle
+ \sum_{k,k',k''} L^{(7)}(j,k,k',k'') m_km_{k'}m_{k''} 
\\& \displaystyle +\sum_{k,k',k''} \sum_{\delta,\delta'} 
L^{(8)}(j,k,k',k'',\delta,\delta'')
\Big(\Delta^{(S)}_\delta(k)\big(\Delta^{(S)}_{\delta'}(k')\big)^*
- \Delta^{(T)}_\delta(k)\big(\Delta^{(T)}_{\delta'}(k')\big)^*\Big)m_{k''}, 
\end{array}\end{equation}
where the functions $L^{(n)}$ $(n=1, \cdots, 8)$  are given in Appendix A.  

From eqs. (11)-(13),  it is seen that the equations for 
$\Delta^{(S)}$, $\Delta^{(T)}$, and $m$ have 
the second-order terms of the forms $m \Delta^{(T)}$, $m \Delta^{(S)}$, 
and $\Delta^{(S)} \Delta^{(T)}$, respectively. 
This implies that the GL free energy should have the cubic term of the 
form $m \Delta^{(S)} \Delta^{(T)}$, and it is actually the case as 
we will see in the following sections. 
It should be noted that eqs. (11)-(13) are valid even when the OPs
have rapid spatial variations, because we have not yet taken a continuum limit. 
This property is important when we consider the antiferromagnetic case 
in $\S$4.

\section{GL Equations for Coexistent States of 
Superconductivity and Ferromagnetism}

In this section, we consider the coexistent states of superconductivity 
and ferromagnetism. 
The GL equation for the SCOP of each symmetry can be obtained by making a 
linear combination of eqs. (11) and (12):   
\begin{equation}\begin{array}{rl}
 \Delta_s(j) = &  \displaystyle \frac{1}{4} \sum_{\eta=\pm{\hat x},\pm{\hat y}}
 \Delta_\eta^{(S)}(j) ,  \ \ 
 \Delta_d(j) =  \frac{1}{4} \big[\sum_{\eta=\pm{\hat x}}  \Delta_\eta^{(S)}(j) 
 - \sum_{\eta=\pm{\hat y}}  \Delta_\eta^{(S)}(j)\big],  
 \\
\Delta_{px}(j) = & \displaystyle \frac{1}{2} \big[\Delta_{\hat x}^{(T)}(j) 
- \Delta_{-{\hat x}}^{(T)}(j) \big], \ \ 
 \Delta_{py}(j) = \frac{1}{2}  \big[\Delta_{\hat y}^{(T)}(j) 
- \Delta_{-{\hat y}}^{(T)}(j) \big]. \\
\end{array}\end{equation}
Assuming that the SCOPs and  magnetization are slowly varying,  
we take a continuum limit.  The SCOPs in the linear and quadratic terms are 
expanded in powers of derivatives by denoting  
${\bf r}_j \to {\bf r}$, ${\bf r}_k\to {\bf r}'$: 
\begin{equation}\begin{array}{rl}
\displaystyle \Delta_\delta(k) & \to \Delta_\delta({\bf r}') 
\\ & \displaystyle 
\sim \Delta_\delta({\bf r}) 
+ ({\bf r}'-{\bf r})_\mu \nabla_\mu \Delta_\delta({\bf r})  
+ \frac{1}{2 } ({\bf r}'-{\bf r})_\mu ({\bf r}'-{\bf r})_\nu \nabla_\mu \nabla_\nu
 \Delta_\delta({\bf r}), \\
\end{array}\end{equation}
where the summations over $\mu$ and $\nu$ are assumed, and a similar 
approximation is carried out for $m$.   
The Peierls phase coming from ${\tilde G}_0$  is also expanded in powers 
of ${\bf  A}$.  Using  the approximation 
%
$
\phi_{k,j} \sim  
 -\frac{\pi}{\phi_0} ({\bf r}'-{\bf r})\cdot{\bf A}({\bf r})$,  
%
the derivatives and ${\bf A}$ are combined to construct 
the gauge-invariant gradient acting on $\Delta$, 
$\displaystyle {\bf D} \equiv {\bf \nabla} +\frac{2\pi i}{\phi_0}{\bf A}$,  
and we keep the terms up to the second order in ${\bf D}$. 
As a typical example for treating  the derivative terms,  
the derivation of the second-order term for $\Delta_s$ is presented  
in Appendix B. 
%
In the third-order terms, we neglect the derivative terms and 
the vector potential ${\bf A}$ as usual, 
namely, $\Delta^{(S)}$, $\Delta^{(T)}$, and $m$ with the arguments 
$k$, $k'$, and $k^{''}$ are replaced with $\Delta^{(S)}({\bf r})$,  
$\Delta^{(T)}({\bf r})$, and $m({\bf r})$, respectively. 
Rewriting $\Delta^{(S)}$ ($\Delta^{(T)}$) using 
$\Delta_s$ and $\Delta_d$ ($\Delta_{px}$ and $\Delta_{py}$),   
we carry out straightforward but lengthy calculations to get 
the following GL equations for SCOPs 
and magnetization: 
\begin{equation}\begin{array}{rl}
& \displaystyle 
\alpha_s \Delta_s + 2\beta_s |\Delta_s|^2\Delta_s 
- K_s (D_x^2+D_y^2) \Delta_s - K_{sd}(D_x^2-D_y^2)\Delta_d \\
+  & \displaystyle K_{spm}\Big[(\nabla_x m)\Delta_{px} + (\nabla_y m)\Delta_{py} 
+ 2m(D_x\Delta_{px}+D_y\Delta_{py})\Big] \\
+  & \displaystyle \gamma_1|\Delta_d|^2\Delta_s + 2\gamma_2\Delta_d^2\Delta_s^*
+ \gamma_3(|\Delta_{px}|^2+|\Delta_{py}|^2)\Delta_s 
+ 2\gamma_5(\Delta_{px}^2+\Delta_{py}^2)\Delta_s^* \\
+  & \displaystyle \gamma_7(|\Delta_{px}|^2-|\Delta_{py}|^2)\Delta_d
+ \gamma_8(\Delta_{px}^2-\Delta_{py}^2)\Delta_d^* 
+ \gamma_{ms}m^2\Delta_s = 0, 
\end{array}\end{equation}
\begin{equation}\begin{array}{rl}
& \displaystyle 
\alpha_d \Delta_d + 2\beta_d |\Delta_d|^2\Delta_d 
- K_d (D_x^2+D_y^2) \Delta_d - K_{sd}(D_x^2-D_y^2)\Delta_s \\
+  & \displaystyle K_{dpm}\Big[(\nabla_x m)\Delta_{px} - (\nabla_y m)\Delta_{py} 
+ 2m(D_x\Delta_{px}-D_y\Delta_{py})\Big] \\
+  & \displaystyle \gamma_1|\Delta_s|^2\Delta_d + 2\gamma_2\Delta_s^2\Delta_d^*
+ \gamma_4(|\Delta_{px}|^2+|\Delta_{py}|^2)\Delta_d 
+ 2\gamma_6(\Delta_{px}^2+\Delta_{py}^2)\Delta_d^* \\
+  & \displaystyle \gamma_7(|\Delta_{px}|^2-|\Delta_{py}|^2)\Delta_s 
+ \gamma_8(\Delta_{px}^2-\Delta_{py}^2)\Delta_s^* 
+ \gamma_{md}m^2\Delta_d = 0, 
\end{array}\end{equation}
\begin{equation}\begin{array}{rl}
& \displaystyle 
\alpha_p \Delta_{px(y)} + 2\beta_p |\Delta_{px(y)}|^2\Delta_{px(y)} 
-K_{p1}D_{x(y)}^2\Delta_{px(y)}-K_{p2}D_{y(x)}^2\Delta_{px(y)} \\
- & \displaystyle (K_{p3}+K_{p4})D_xD_y\Delta_{py(x)} 
- K_{spm}\Big[(\nabla_{x(y)}m)\Delta_s+2mD_{x(y)}\Delta_s\Big] \\
- & \displaystyle K_{dpm}\Big[(\nabla_{x(y)}m)\Delta_d+2mD_{x(y)}\Delta_d\Big] 
+  \gamma_{p1}|\Delta_{py(x)}|^2\Delta_{px(y)} 
+ 2\gamma_{p2}\Delta_{py(x)}^2\Delta_{px(y)}^*  \\
+ & \displaystyle \gamma_3|\Delta_s|^2\Delta_{px(y)} 
+ \gamma_4|\Delta_d|^2\Delta_{px(y)} 
+  2\gamma_5 \Delta_s^2\Delta_{px(y)}^*
 + 2\gamma_6\Delta_d^2\Delta_{px(y)}^*  \\
\pm & \displaystyle \gamma_7(\Delta_s\Delta_d^*+c.c.)\Delta_{px(y)} 
\pm 2\gamma_8 \Delta_s\Delta_d\Delta_{px(y)}^* 
+ \gamma_{mp} m^2\Delta_{px(y)} = 0,  
\end{array}\end{equation}
\begin{equation}\begin{array}{rl}
& \displaystyle 
\alpha_m m+ 2\beta_m m^3 -K_m(\nabla_x^2+\nabla_y^2)m \\
+ & \displaystyle \frac{1}{2} K_{spm} \Big[\Delta_s\{(D_x\Delta_{px})^* 
+ (D_y\Delta_{py})^*\} - \{\Delta_{px}^*D_x\Delta_s 
+ \Delta_{py}^*D_y\Delta_s\}+c.c.\Big] \\ 
+ & \displaystyle \frac{1}{2} K_{dpm} \Big[\Delta_d\{(D_x\Delta_{px})^* 
- (D_y\Delta_{py})^*\} - \{\Delta_{px}^*D_x\Delta_d 
- \Delta_{py}^*D_y\Delta_d\}+c.c.\Big] \\ 
+ & \displaystyle \gamma_{ms}m|\Delta_s|^2 + \gamma_{md}m|\Delta_d|^2 
+ \gamma_{mp}m(|\Delta_{px}|^2+|\Delta_{py}|^2) = 0,
\end{array}\end{equation}
where the coefficients appearing in eqs. (16)-(19) are 
given in Appendix C.

Equations (16)-(19) are the coupled equations that determine 
the SCOPs and  magnetization self-consistently. 
The most important point is that the second-order terms with a first-order 
derivative exist in the GL equations. They can induce triplet (singlet) SCOPs 
in a singlet (triplet) superconductor once the magnetization coexists 
inhomogeneously.  
It should also be noted that the coefficients in GL equations 
are determined microscopically, reflecting the nature of the 
electronic states of the original model, {\it e.g.}, the shape of the 
Fermi surface. This property can be used to study the 
coexistent states of realistic materials to be considered.

The GL free energy $F$ up to the fourth order in OPs 
can be obtained from the above GL equations in such a way that the 
variations of $F$ with respect to OPs reproduce eqs. (16)-(19).  
The results are written as follows: 
\begin{equation}\begin{array}{rl}
 \displaystyle F = F_S  + & \displaystyle F_T + F_{ST} + F_M + F_{SM} + F_{TM} 
 + F_{STM}, \\
F_S = &\displaystyle  \int d^2{\bf r} \Big[
 \alpha_s |\Delta_s|^2 + \beta_s |\Delta_s|^4  + K_s |{\vec D} \Delta_s|^2 
 + \alpha_d |\Delta_d|^2+ \beta_d |\Delta_d|^4 +  K_d |{\vec D} \Delta_d|^2 \\
 & \displaystyle + \gamma_1 |\Delta_s|^2|\Delta_d|^2 
 + \gamma_2 \big(\Delta_d^2(\Delta_s^*)^2 + c.c.\big)  \\
& + K_{ds} \big((D_x\Delta_d)(D_x\Delta_s)^{*} 
 - (D_y\Delta_d)(D_y\Delta_s)^{*} + c.c. \big)\Big],   \\
 F_T =   &\displaystyle \int d^2{\bf r} \Big[
\alpha_p\big(|\Delta_{px}|^2 + |\Delta_{py}|^2\big) 
+ \beta_p\big(|\Delta_{px}|^4 + |\Delta_{py}|^4\big)  \\
& \displaystyle + \gamma_{p1}|\Delta_{px}|^2|\Delta_{py}|^2 
+ \gamma_{p2}\big(\Delta_{px}^2(\Delta_{py}^{*})^2 + c.c.\big) \\
& \displaystyle + K_{p1}\big(|D_x\Delta_{px}|^2 + |D_y\Delta_{py}|^2\big)
+ K_{p2}\big(|D_y\Delta_{px}|^2 + |D_x\Delta_{py}|^2\big) \\
& \displaystyle + K_{p3}\big((D_x\Delta_{px})^{*}(D_y\Delta_{py})+ c.c.\big)
+ K_{p4}\big((D_y\Delta_{px})^{*}(D_x\Delta_{py})+ c.c.\big)\Big], \\
 F_{ST} = &\displaystyle  \int d^2{\bf r} \Big[
 \gamma_3
 \big(|\Delta_{px}|^2 + |\Delta_{py}|^2\big)|\Delta_s|^2 
+ \gamma_4 \big(|\Delta_{px}|^2 + |\Delta_{py}|^2\big)|\Delta_d|^2 \\
& \displaystyle + \gamma_5 
\big\{\big(\Delta_{px}^2 + \Delta_{py}^2\big)(\Delta_s^{*})^2 + c.c.\big\}
+ \gamma_6 
\big\{\big(\Delta_{px}^2 + \Delta_{py}^2\big)(\Delta_d^{*})^2 + c.c.\big\} \\
& \displaystyle + \gamma_7 
\big(|\Delta_{px}|^2 - |\Delta_{py}|^2\big)\big(\Delta_s^{*}\Delta_d + c.c.\big) 
+ \gamma_8 
\big\{\big(\Delta_{px}^2 - \Delta_{py}^2\big)\Delta_s^{*}\Delta_d^{*} 
+ c.c.\big\}\Big], \\
 F_M =  &\displaystyle \int d^2{\bf r} \Big[
\alpha_m  m^2 + \beta_m m^4 + K_m \big(\nabla m\big)^2\big], \\ 
 F_{SM}  = &\displaystyle \int d^2{\bf r} \Big(
\gamma_{ms} m^2 |\Delta_s|^2 + \gamma_{md} m^2 |\Delta_d|^2 \Big), \\
F_{TM}  = &\displaystyle \int d^2{\bf r} 
\gamma_{mp} m^2\Big(|\Delta_{px}|^2 + |\Delta_{py}|^2\Big), \\
\displaystyle F_{STM}  = & \int d^2{\bf r}  \displaystyle  \Big\{
K_{spm} m \Big[\Delta_s\Big((D_x\Delta_{px})^*+(D_y\Delta_{py})^*\Big)  
- \Big((D_x\Delta_s)\Delta_{px}^* 
 +(D_y\Delta_s)\Delta_{py}^*\Big)\Big] \\
& \displaystyle + K_{dpm} m 
\Big[\Delta_d\Big((D_x\Delta_{px})^*-(D_y\Delta_{py})^*\Big)  
- \Big((D_x\Delta_d)\Delta_{px}^*-(D_y\Delta_d)\Delta_{py}^*\Big)\Big]
+ c.c.\Big\}.   \\
\end{array}\end{equation}
Here, $F_S$, $F_T$, and $F_M$ are the free energy for the singlet and 
triplet SCOPs and 
the magnetization, respectively, while $F_{ST}$, $F_{SM}$, $F_{TM}$, and 
$F_{STM}$ describe their couplings. 
Note that $F$ is invariant under all the symmetry operations of the square 
lattice.\cite{SigUed}
The cubic term $F_{STM}$ has derivative couplings of singlet and 
triplet SCOPs with magnetization, so that triplet (singlet) SCOPs 
would be induced once ferromagnetism coexists with singlet (triplet) 
superconductivity inhomogeneously, as already noted.
(In other words, the triplet (singlet) SCOP would not be induced if the 
coexistence occurs uniformly.) 
This gives a simple and clear interpretation for previous 
theoretical results using the BdG method\cite{KK1} 
and quasiclassical Green's functions,\cite{Berg2,Esch} 
in which the occurrence of $p$-wave SCOPs near the interface between 
a singlet superconductor and a ferromagnet was pointed out.

Dahl and  Sudb\o\cite{Dahl} derived the GL free energy from a model 
with a spin generalized BCS term and a Heisenberg exchange term, 
which is different from ours. 
They found a cubic term in the GL free energy 
that couples a nonunitary SCOP with magnetization.

\section{Case of Superconductivity and Antiferromagnetism}
Next, we consider the coexistent states of superconductivity and 
antiferromagnetism.  
In the AF state the magnetization 
$m_j$ is oscillating, so we expect that the triplet SCOP will be induced 
even in a uniform AF state once the coexistence occurs. 
As a  slowly varying OP to be considered in the continuum theory, 
we define the staggered magnetization   
$M_j \equiv \displaystyle m_j e^{i{\bf Q}\cdot{\bf r}_j}$ with 
 $ {\bf Q} \equiv (\pi,\pi)$. 
If we assume that the singlet component of SCOP,  
$\Delta^{(S)}$, is also slowly varying, then the triplet component 
$\Delta^{(T)}$ should oscillate, as can be  seen from eqs. (11)-(13).  
Therefore,  we define the $\pi$-triplet SCOP 
$\displaystyle \Delta^{(\pi T)}_\eta(j) \equiv  
\Delta^{(T)}_\eta(j) e^{i{\bf Q}\cdot{\bf r}_j}$.     
Rewriting eqs. (11)-(13) in terms of $M$ and $\Delta^{(\pi T)}$, 
we find that all terms 
in these equations do not have staggered oscillations. 
Defining the $p_x$ and $p_y$ components of the $\pi$-triplet SCOP as 
\begin{equation}\begin{array}{rl}
\displaystyle \Delta^{(\pi T)}_{px}(j) = & \displaystyle
\frac{1}{2} \big[\Delta^{(\pi T)}_{\hat x}(j) + \Delta^{(\pi T)}_{-{\hat x}}(j)\big], 
\\ 
\displaystyle \Delta^{(\pi T)}_{py}(j) = & \displaystyle
\frac{1}{2} \big[\Delta^{(\pi T)}_{\hat y}(j) + \Delta^{(\pi T)}_{-{\hat y}}(j)\big], 
\end{array}\end{equation}
we carry out calculations similar to those in the ferromagnetic case to get  GL 
equations and the GL free energy. 
Here, we present only the resulting expressions for the free energy  
$F^{AF}$: 
\begin{equation}\begin{array}{rl}
 \displaystyle F^{AF} = F_S  + & \displaystyle F_T^{AF} + F_{ST}^{AF} 
 + F_M^{AF} + F_{SM}^{AF} + F_{TM}^{AF} + F_{STM}^{AF}, \\
 F_T^{AF} =   &\displaystyle \int d^2{\bf r} \Big[
{\tilde \alpha}_{p1}\big(|\Delta_{px}^{(\pi T)}|^2 + |\Delta_{py}^{(\pi T)}|^2\big) 
+ {\tilde \alpha}_{p2}\big(\Delta_{px}^{(\pi T)}(\Delta_{py}^{(\pi T)})^* + c.c\big) 
+ {\tilde \beta}_p\big(|\Delta_{px}|^4 + |\Delta_{py}|^4\big)  \\
& \displaystyle + {\tilde \gamma}_{p1}
|\Delta_{px}^{(\pi T)}|^2|\Delta_{py}^{(\pi T)}|^2  
+ {\tilde \gamma}_{p2}\big((\Delta_{px}^{(\pi T)})^2(\Delta_{py}^{(\pi T)*})^2 
+ c.c.\big) \\
& \displaystyle 
+ {\tilde \gamma}_{p3}
\big(|\Delta_{px}^{(\pi T)}|^2 + |\Delta_{py}^{(\pi T)}|^2\big) 
\big(\Delta_{px}^{(\pi T)}(\Delta_{py}^{(\pi T)})^*+ c.c.\big) \\
& \displaystyle + {\tilde K}_{p1}\big(|D_x\Delta_{px}^{(\pi T)}|^2 
+ |D_y\Delta_{py}^{(\pi T)}|^2\big)
+ {\tilde K}_{p2}\big(|D_y\Delta_{px}^{(\pi T)}|^2 
+ |D_x\Delta_{py}^{(\pi T)}|^2\big) \\
& \displaystyle + {\tilde K}_{p3}\big((D_x\Delta_{px}^{(\pi T)})^{*}
(D_x\Delta_{py}^{(\pi T)})
+ (D_y\Delta_{px}^{(\pi T)})^{*}(D_y\Delta_{py}^{(\pi T)}) + c.c.\big)\Big], \\

 F_{ST}^{AF} = &\displaystyle  \int d^2{\bf r} \Big[
 {\tilde \gamma}_3
 \big(|\Delta_{px}^{(\pi T)}|^2 + |\Delta_{py}^{(\pi T)}|^2\big)|\Delta_s|^2 
+ {\tilde \gamma}_4 \big(|\Delta_{px}^{(\pi T)}|^2 
+ |\Delta_{py}^{(\pi T)}|^2\big)|\Delta_d|^2 \\
& \displaystyle + {\tilde \gamma}_5 
\big\{\big((\Delta_{px}^{(\pi T)})^2 + (\Delta_{py}^{(\pi T)})^2\big)
(\Delta_s^{*})^2 + c.c.\big\}
+ {\tilde \gamma}_6 
\big\{\big((\Delta_{px}^{(\pi T)})^2 + (\Delta_{py}^{(\pi T)})^2\big)
(\Delta_d^{*})^2 + c.c.\big\} \\
& \displaystyle + {\tilde \gamma}_7 
\big(|\Delta_{px}^{(\pi T)}|^2 - |\Delta_{py}^{(\pi T)}|^2\big)
\big(\Delta_s^{*}\Delta_d + c.c.\big) 
+ {\tilde \gamma}_8 
\big\{\big((\Delta_{px}^{(\pi T)})^2 - (\Delta_{py}^{(\pi T)})^2\big)
\Delta_s^{*}\Delta_d^{*} + c.c.\big\}\Big] \\
& \displaystyle + {\tilde \gamma}_9 
\big((\Delta_{px}^{(\pi T)})^*\Delta_{py}^{(\pi T)}+c.c.\big) |\Delta_s|^2
+  {\tilde \gamma}_{10}
\big((\Delta_{px}^{(\pi T)})^*\Delta_{py}^{(\pi T)}+c.c.\big)  |\Delta_d|^2\\  
& \displaystyle + {\tilde \gamma}_{11} 
\big(\Delta_{px}^{(\pi T)}\Delta_{py}^{(\pi T)}(\Delta_s^*)^2+c.c.) 
+ {\tilde \gamma}_{12} 
\big(\Delta_{px}^{(\pi T)}\Delta_{py}^{(\pi T)}(\Delta_d^*)^2+c.c.)\Big], \\
 F_M^{AF} =  &\displaystyle \int d^2{\bf r} \Big[
{\tilde \alpha}_m  M^2 + {\tilde \beta}_m M^4 
+ {\tilde K}_m \big(\nabla M\big)^2\big], \\ 
 F_{SM}^{AF}  = &\displaystyle \int d^2{\bf r} \Big(
{\tilde \gamma}_{ms} M^2 |\Delta_s|^2 
+ {\tilde \gamma}_{md} M^2 |\Delta_d|^2 \Big), \\
F_{TM}^{AF}  = &\displaystyle \int d^2{\bf r} \Big[ 
{\tilde \gamma}_{mp1} M^2\Big(|\Delta_{px}^{(\pi T)}|^2  
+ |\Delta_{py}^{(\pi T)}|^2\Big) 
+ {\tilde \gamma}_{mp2} M^2 \Big(\Delta_{px}^{(\pi T)}
(\Delta_{py}^{(\pi T)})^* + c.c.\Big)\Big],  \\
\displaystyle F^{AF}_{STM}  =  &\displaystyle \int d^2{\bf  r}  \Big[
{\tilde \gamma}_{spm} M \Delta_s\big(\Delta_{px}^{(\pi T)} 
+ \Delta_{py}^{(\pi T)}\big)^*  
+ {\tilde \gamma}_{dpm} M \Delta_d\big(\Delta_{px}^{(\pi T)} 
- \Delta_{py}^{(\pi T)}\big)^* 
+ c.c.\Big], 
\end{array}\end{equation}
where $F_S$ is the same as that in the ferromagnetic case. 
The expressions for the coefficients appearing in $F^{AF}$ are summarized in Appendix D.
The cubic term $F^{AF}_{STM}$ in this case 
couples $\Delta^{(S)}$, $\Delta^{(\pi T)}$, and $M$ directly without 
derivatives. Then the $(p_x+p_y)$-wave  [$(p_x-p_y)$-wave]   
$\pi$-triplet component would  be induced when $s$-wave ($d$-wave)
superconductivity and antiferromagnetism coexist, even in a uniform case. 
This is consistent with the results of previous mean-field calculations 
that  predict the occurrence of the $\pi$-triplet component in uniformly coexistent 
states of $d$-wave superconductivity and 
antiferromagnetism.\cite{Fenton,Mura1,Mura2,Kyung,Apens}  
This term also gives a simple explanation for 
the occurrence of the $\pi$-triplet SCOP near  
the interface between a singlet superconductor and an antiferromagnet, 
which was found in numerical calculations based on the BdG method.\cite{KK1}

\section{Summary and Discussion}
We have derived  GL equations and the GL free energy for the 
coexistent states of superconductivity and magnetism microscopically 
from the extended Hubbard model with on-site repulsive and nearest-neighbor 
attractive interactions. 
It was found that, in the GL free energy, a cubic term that couples
singlet and triplet SCOPs with magnetization exists. 
Owing to this term, triplet SCOPs would be induced when ferromagnetism coexists
with singlet superconductivity inhomogeneously. 
This gives a simple explanation for previous theoretical studies on bilayer 
systems composed of a ferromagnet and a singlet 
superconductor.\cite{KK1,Berg2,Esch} 
 In the coexistent state of antiferromagnetism and singlet superconductivity,  
$\pi$-triplet SCOPs would be induced. This occurs  not only in inhomogeneous 
cases but also in spatially uniform states.

The validity of the model employed in this paper is limited because 
of the absence of the $SU(2)$ symmetry in spin space. 
For more general and precise argument of the symmetry of the 
induced OPs, theoretical investigations based 
on the model that respects this symmetry will be necessary, 
although the present study may capture some of the important aspects.

In order to study the material dependence of interface states  
more generally, it is necessary to derive  GL equations from other 
microscopic models.   
For example, the low-energy electronic states of high-$T_C$ 
cuprate superconductors are described by the $t-J$ model,\cite{Ogata} 
and so the interface state of heterostructures made of high-$T_C$ cuprates 
and magnetic materials may be studied using the GL equations derived from 
this model. 

Numerical study of the GL equations derived from different microscopic 
models may clarify the material dependence of the interface states of 
heterostructures composed of various superconductors and magnetic materials. 
This problem will be examined separately.

\begin{acknowledgment}
K. K. thanks H. Yamase for useful discussions. 

\end{acknowledgment}

\appendix
\section{Functions Appearing in GL Equations}
The functions $L^{(n)}$ $(n=1, \cdots, 8)$ appearing in eqs. (12)-(14) 
are defined as follows:  
\begin{equation}\begin{array}{rl}
L^{(1)}(j,k,\eta,\delta) = & \displaystyle 
VT\sum_{\varepsilon_n} \sum_{k,\delta} 
{\tilde G}_0(k,j+\eta,-i\varepsilon_n){\tilde G}_0(k+\delta,j,i\varepsilon_n), 
\\ 
L^{(2)}(j,k,k',\eta,\delta) = & \displaystyle 
VUT\sum_{\varepsilon_n} 
\Big[{\tilde G}_0(k,j+\eta,-i\varepsilon_n){\tilde G}_0(k+\delta,k',i\varepsilon_n)
{\tilde G}_0(k',j,i\varepsilon_n) \\
& \displaystyle 
-{\tilde G}_0(k',j+\eta,-i\varepsilon_n){\tilde G}_0(k,k',-i\varepsilon_n)
{\tilde G}_0(k+\delta,j,i\varepsilon_n)\Big],  
\\
L^{(3)}(j,k,k',k'',\eta,\delta.\delta',\delta'') = & \displaystyle 
- V^3T\sum_{\varepsilon_n} 
 {\tilde G}_0(k,j+\eta,-i\varepsilon_n) {\tilde G}_0(k+\delta,k',i\varepsilon_n)
\\  & \displaystyle
 \times {\tilde G}_0(k'',k'+\delta',-i\varepsilon_n){\tilde G}_0(k''+\delta'',j,i\varepsilon_n),  
\\ 
L^{(4)}(j,k,k',k'',\eta,\delta) = & \displaystyle 
VU^2T\sum_{\varepsilon_n}
 \Big[{\tilde G}_0(k,j+\eta,-i\varepsilon_n)
{\tilde G}_0(k+\delta,k',i\varepsilon_n) 
\\ & \displaystyle 
\times {\tilde G}_0(k',k'',i\varepsilon_n)  {\tilde G}_0(k'',j,i\varepsilon_n)
\\ & \displaystyle 
- {\tilde G}_0(k',j+\eta,-i\varepsilon_n)  {\tilde G}_0(k,k',-i\varepsilon_n)
{\tilde G}_0(k+\delta,k'',i\varepsilon_n)
{\tilde G}_0(k'',j,i\varepsilon_n) 
\\& \displaystyle 
+ {\tilde G}_0(k'',j+\eta,-i\varepsilon_n)
{\tilde G}_0(k',k'',-i\varepsilon_n) 
{\tilde G}_0(k,k',-i\varepsilon_n){\tilde G}_0(k+\delta,j,i\varepsilon_n)\Big], 
\\ 
L^{(5)}(j,k) = & \displaystyle 
 -UT\sum_{\varepsilon_n} 
 {\tilde G}_0(j,k,i\varepsilon_n){\tilde G}_0(k,j,i\varepsilon_n), 
\\ 
L^{(6)}(j,k,k',\delta,\delta') = & \displaystyle 
V^2T\sum_{\varepsilon_n} 
 {\tilde G}_0(j,k,i\varepsilon_n)
 {\tilde G}_0(k',k+\delta,-i\varepsilon_n)
 {\tilde G}_0(k'+\delta',j,i\varepsilon_n), 
\\ 
L^{(7)}(j,k,k',k'') = & \displaystyle 
- U^3T\sum_{\varepsilon_n} 
{\tilde G}_0(j,k,i\varepsilon_n){\tilde G}_0(k,k',i\varepsilon_n)
{\tilde G}_0(k',k'',i\varepsilon_n)
{\tilde G}_0(k'',j,i\varepsilon_n), 
\\ 
L^{(8)}(j,k,k',k'',\delta,\delta') = & \displaystyle 
V^2UT\sum_{\varepsilon_n}
\Big[{\tilde G}_0(j,k,i\varepsilon_n)
{\tilde G}_0(k',k+\delta,-i\varepsilon_n) 
\\ & \displaystyle 
\times {\tilde G}_0(k'+\delta',k'',i\varepsilon_n){\tilde G}_0(k'',j,i\varepsilon_n) \\
& \displaystyle + {\tilde G}_0(j,k'',i\varepsilon_n){\tilde G}_0(k'',k,i\varepsilon_n)
{\tilde G}_0(k',k+\delta,-i\varepsilon_n){\tilde G}_0(k'+\delta',j,i\varepsilon_n) \\
& \displaystyle - {\tilde G}_0(j,k,i\varepsilon_n)
{\tilde G}_0(k'',k+\delta,-i\varepsilon_n)
{\tilde G}_0(k',k'',-i\varepsilon_n){\tilde G}_0(k'+\delta',j,i\varepsilon_n)\Big]. 
\\ 
\end{array}\end{equation}

\appendix
\section{Derivation of the Second Order Terms in GL Equations}
In this appendix, we show how to calculate the second-order terms in the GL 
equations for ferromagnetism and superconductivity. 
Here, the equation for $\Delta_s$ is treated as an example. 
(Other OPs can be treated similarly.)  
The term to be considered is 
\begin{equation}\begin{array}{rl} 
 \displaystyle \frac{1}{4} 
 \sum_\eta VU T\sum_{\varepsilon_n} \sum_{k,k',\delta} 
& \displaystyle  \big (\Delta_\delta^{(T)}(k)\big)^* m_{k'}
 \Big[{\tilde G}_0(k,j+\eta,-i\varepsilon_n){\tilde G}_0(k+\delta,k',i\varepsilon_n)
{\tilde G}_0(k',j,i\varepsilon_n) \\
& \displaystyle 
-{\tilde G}_0(k',j+\eta,-i\varepsilon_n){\tilde G}_0(k,k',-i\varepsilon_n)
{\tilde G}_0(k+\delta,j,i\varepsilon_n)\Big].  
\end{array}\end{equation}
We substitute eq. (15) for $\Delta^{(T)}_\delta(k)$ and use 
a similar approximation to $m_{k'}$,  
and denote ${\bf r}_j \to {\bf r}$, ${\bf r}_k \to {\bf r}'$ and 
${\bf r}_{k'} \to {\bf r}^{''}$. 
The term without derivatives and ${\bf A}$ is given by 
\begin{equation}\begin{array}{rl}  
\displaystyle \frac{1}{4}VUm({\bf r})
& \displaystyle  T\sum_{\varepsilon_n} \frac{1}{N}\sum_p 
\sum_\eta e^{-i{\bf p}\cdot{\bf \eta}} 
\sum_\delta \big (\Delta_\delta^{(T)}({\bf r})\big)^* 
 e^{-i{\bf p}\cdot{\bf \delta}}
\big[G_0^2({\bf p},i\varepsilon_n)G_0({\bf p},-i\varepsilon_n) \\
& \displaystyle 
- G_0({\bf p},i\varepsilon_n)G_0^2({\bf p},-i\varepsilon_n)\big], 
\end{array}\end{equation} 
where $N$ is the total number of lattice sites. 
This term is seen to vanish by putting $\varepsilon_n \to -\varepsilon_n$ 
in the second line. 
Next  the terms that  are first order in derivatives and ${\bf A}$ are given as  
 \begin{equation}\begin{array}{rl} 
\displaystyle \frac{1}{4}VU & \displaystyle 
\sum_\eta T\sum_{\varepsilon_n} \sum_\delta \int d^2{\bf r}' \int d^2{\bf r}^{''}  
\Big[m({\bf r})\big({\bf r}'-{\bf r}\big)_\mu 
\big\{\nabla_\mu  -\frac{2\pi i}{\phi_0} {\bf A}_\mu({\bf r})\big\}
\Delta^{(T)}_\delta({\bf r}) \\
 & \displaystyle 
+ \Delta^{(T)}_\delta({\bf r})\big({\bf r}^{''}-{\bf r}\big)_\mu \nabla_\mu m({\bf r}) 
\Big]  \frac{1}{N^3}\sum_{p_1 p_2 p_3}
G_0(p_1,-i\varepsilon_n)G_0(p_2,i\varepsilon_n)G_0(p_3,i\varepsilon_n) \\
& \displaystyle \times
e^{i({\bf p}_1+{\bf p}_2)\cdot{\bf r}'}
e^{i({\bf p}_3-{\bf p}_2)\cdot{\bf r}^{''}}
e^{-i({\bf p}_1+{\bf p}_3)\cdot{\bf r}}
\Big(e^{-i{\bf p}_1\cdot{\bf \eta}}e^{i{\bf p}_2\cdot{\bf \delta}}
-e^{i{\bf p}_1\cdot{\bf \delta}}e^{-i{\bf p}_3\cdot{\bf \eta}}\Big). 
\end{array}\end{equation}
Substituting the relations
\begin{equation}\begin{array}{rl}
\displaystyle ({\bf r}'-{\bf r})_\mu e^{i({\bf p}_1+{\bf p}_2)\cdot{\bf r}'}
=  & \displaystyle e^{i({\bf p}_1+{\bf p}_2)\cdot{\bf r}}
\Big(-i\frac{\partial}{\partial p_{2\mu}}\Big)
e^{i({\bf p}_1+{\bf p}_2)\cdot({\bf r}'-{\bf r})}, \\
 \displaystyle ({\bf r}^{''}-{\bf r})_\mu e^{i({\bf p}_3-{\bf p}_2)\cdot{\bf r}^{''}}
=  & \displaystyle e^{i({\bf p}_3-{\bf p}_2)\cdot{\bf r}}
\Big(-i\frac{\partial}{\partial p_{3\mu}}\Big)
e^{i({\bf p}_3-{\bf p}_2)\cdot({\bf r}^{''}-{\bf r})}, 
\end{array}\end{equation}
we carry out the integrations over ${\bf r}'$ and ${\bf r}^{''}$ after 
performing the partial integration on $p_\mu$. 
Then eq. (B$\cdot$3) becomes 
\begin{equation}\begin{array}{rl}
&  \displaystyle -\frac{1}{4}VU \sum_\eta 
T\sum_{\varepsilon_n} \sum_\delta
\frac{1}{N}\sum_p \Big[
\big(\Delta^{(T)}_\delta({\bf r})\big)^*\nabla_\mu m({\bf r}) 
G_0(p,-i\varepsilon_n)G_0(p,i\varepsilon_n)
i\frac{\partial}{\partial p_\mu}G_0(p,i\varepsilon_n)
\\ & \displaystyle \ \ \ \ \ 
+ m({\bf r})\big(D_\mu \Delta^{(T)}_\delta({\bf r})\big)^* 
G_0(p,-i\varepsilon_n)i\frac{\partial}{\partial p_\mu}G_0^2(p,i\varepsilon_n) 
\Big] 
\Big(e^{-i{\bf p}\cdot{\bf \eta}}e^{-i{\bf p}\cdot{\bf \delta}}
- e^{i{\bf p}\cdot{\bf \eta}}e^{i{\bf p}\cdot{\bf \delta}}\Big) \\
&  \sim  \displaystyle -\frac{1}{4}VU \frac{1}{N}\sum_p I_3(p) 
\frac{\partial \xi_p}{\partial p_\mu} \sum_\eta\sum_\delta 
i \Big(e^{-i{\bf p}\cdot{\bf \eta}}e^{-i{\bf p}\cdot{\bf \delta}}
- e^{i{\bf p}\cdot{\bf \eta}}e^{i{\bf p}\cdot{\bf \delta}}\Big) 
\\ & \displaystyle  \ \ \ \ \ \times 
\Big[
\big(\Delta^{(T)}_\delta({\bf r})\big)^*\nabla_\mu m({\bf r})
+ 2m({\bf r}) \big(D_\mu\Delta^{(T)}_\delta({\bf r})\big)^* 
\Big] 
\\ & = \displaystyle 
-2VU \frac{1}{N} \sum_p 
I_3(p)\frac{\partial \xi_p}{\partial p_\mu}
\omega_s \Big[
 (\nabla_\mu m({\bf r}))\big(\Delta_{px}^*({\bf r})\omega_x 
+ \Delta_{py}^*({\bf r})\omega_y\big)
 \\ & \displaystyle \ \ \ \ \ 
+ 2m({\bf r})\Big\{\big(D_\mu\Delta_{px}({\bf r})\big)^*\omega_x 
+ \big(D_\mu\Delta_{py}({\bf r})\big)^*\omega_y\Big\} 
\Big].
\end{array}\end{equation}
With the definitions of $K_{spm}$ and the function $I_3(p)$ in Appendix C, 
the last expression is seen to give the second-order term appearing in eq. (16). 

Terms that  are second order in derivatives and ${\bf A}$ can be 
shown to vanish by carrying out similar calculations.

\appendix
\section{Coefficients in GL free energy for ferromagnetism and 
superconductivity}
The coefficients appearing in GL equations [eqs. (16)-(19)] and the GL free 
energy [eq. (20)] are given as follows:  
\begin{equation}\begin{array}{rl}
& \displaystyle\alpha_{s(d)} = 4V \Big(1-\frac{V}{N}\sum_p 
I_1(p) \omega_{s(d)}^2 \Big), \\
& \displaystyle  \beta_{s(d)} = 8V^4\frac{1}{N}\sum_p 
I_2(p) \omega_{s(d)}^4, \\
& \displaystyle  \gamma_1= 32V^4\frac{1}{N}\sum_p 
I_2(p) \omega_s^2 \omega_d^2, 
\ \ \  \gamma_2 = \frac{1}{4} \gamma_1, \\
& \displaystyle  K_{s(d)}= 2V^2 \frac{1}{N} \sum_p I_2(p) 
\Big(\frac{\partial \xi_p}{\partial p_x}\Big)^2 \omega_{s(d)}^2,  \\
& \displaystyle  K_{sd}= 2V^2 \frac{1}{N} \sum_p I_2(p) 
\Big(\frac{\partial \xi_p}{\partial p_x}\Big)^2 \omega_s \omega_d, \\
& \displaystyle  \alpha_p = 2V \Big(1-\frac{2V}{N}\sum_p 
I_1(p) \omega_x^2 \Big), \\
& \displaystyle  \beta_p = 8V^4\frac{1}{N}\sum_p 
I_2(p) \omega_x^4,\\
& \displaystyle  \gamma_{p1} = 32V^4 \frac{1}{N}\sum_p
I_2(p) \omega_x^2 \omega_y^2, 
\ \ \ \gamma_{p2} = \frac{1}{4} \gamma_{p1}, \\
& \displaystyle  K_{p1(2)} = 2V^2 \frac{1}{N}\sum_p
I_2(p) \Big(\frac{\partial \xi_p}{\partial p_x}\Big)^2 \omega_{x(y)}^2, \\
& \displaystyle K_{p3} + K_{p4} = 4V^2 \frac{1}{N}\sum_p
I_2(p) \Big(\frac{\partial \xi_p}{\partial p_x}\Big)
\Big(\frac{\partial \xi_p}{\partial p_y}\Big) \omega_x \omega_y, \\
& \displaystyle  \{\gamma_3, \ \gamma_4, \ \gamma_7\} = 
32V^4 \frac{1}{N}\sum_p I_2(p) \{\omega_s^2\omega_x^2, \ 
\omega_d^2\omega_x^2, \ \omega_s\omega_d\omega_x^2\}, \\
& \displaystyle  \gamma_5 = -\frac{1}{4}\gamma_3, \ 
 \gamma_6 = -\frac{1}{4}\gamma_4, \ \gamma_8 = -\frac{1}{2}\gamma_7, \\
&  \displaystyle  \alpha_m = U\Big(1+\frac{U}{N}\sum_p f^{'}(p)\Big), \\
&  \displaystyle  \beta_m = \frac{U^4}{12N}\sum_p f^{'''}(p),\\
&  \displaystyle  K_m = -\frac{U^2}{12N}\sum_p 
\Big(\frac{\partial \xi_p}{\partial p_x}\Big)^2  f^{'''}(p), \\
&  \displaystyle  \{\gamma_{ms}, \ \gamma_{md}, \ \gamma_{mp}\} 
 = -4V^2U^2 \frac{1}{N} \sum_p(2 I_3(p)-I_2(p)) \{\omega_s^2,  
 \omega_d^2, \omega_x^2\}, \\
 & \displaystyle K_{spm} = 8V^2U \frac{1}{N} \sum_p I_3(p) 
 \frac{\partial \xi_p}{\partial p_x} \omega_s\omega_x, \\
 & \displaystyle K_{dpm} = 8V^2U \frac{1}{N} \sum_p I_3(p) 
 \frac{\partial \xi_p}{\partial p_x} \omega_d\omega_x, 
\end{array}\end{equation}
where $\omega_s= \cos p_x + \cos p_y$, $\omega_d= \cos p_x - \cos p_y$,  
and $\omega_{x(y)}=\sin p_{x(y)}$, and the summation on $p$ is taken 
over the first Brillouin zone.
The functions $I_1$, $I_2$,  and $I_3$ are defined as  
$\displaystyle I_1(p) = 
T\sum_{\varepsilon_n} G_0(p,i\varepsilon_n)G_0(p,-i\varepsilon_n)$, 
$\displaystyle I_2(p) = 
T\sum_{\varepsilon_n} G_0^2(p,i\varepsilon_n)G_0^2(p,-i\varepsilon_n)$,  
and $\displaystyle I_3(p) =  
T\sum_{\varepsilon_n}  G_0^3(p,i\varepsilon_n)G_0(p,-i\varepsilon_n)$,  
with $f(\xi_p)$ being the Fermi distribution function.

\appendix
\section{Coefficients in GL Free Energy for Antiferromagnetism and Superconductivity}
The expressions of the coefficients in $F^{AF}$ [eq. (22)] are given as follows:   
\begin{equation}\begin{array}{rl}
& \displaystyle  {\tilde \alpha}_{p1} = 2V \Big(1-\frac{2V}{N}\sum_p 
I_4(p) \cos^2p_x \Big) , \\
& \displaystyle
{\tilde  \alpha}_{p2} = -\frac{4V^2}{N}\sum_p 
I_4(p) \cos p_x \cos p_y,  \\
& \displaystyle  {\tilde \beta}_p = 8V^4\frac{1}{N}\sum_p 
I_5(p) \cos^4p_x,\\
& \displaystyle  {\tilde \gamma}_{p1} = 32V^4 \frac{1}{N}\sum_p
I_5(p) \cos^2p_x\cos^2p_y, 
\ \ \ {\tilde \gamma}_{p2} = \frac{1}{4} {\tilde \gamma}_{p1}, \\
& \displaystyle  {\tilde \gamma}_{p3} = 16V^4 \frac{1}{N}\sum_p
I_5(p) \cos^3p_x\cos p_y, \\
& \displaystyle  {\tilde K}_{p1(2)} = -2V^2 \frac{1}{N}\sum_p
I_5(p) \Big(\frac{\partial \xi_p}{\partial p_x}\Big)^2 \cos^2p_{x(y)}, \\ 
 & \displaystyle 
{\tilde K}_{p3} = -2V^2 \frac{1}{N}\sum_p
I_5(p) \Big(\frac{\partial \xi_p}{\partial p_x}\Big)^2 \cos p_x\cos p_y, \\
& \displaystyle {\tilde \gamma}_{3(4)} = 32V^4\frac{1}{N} \sum_p I_6(p) 
\omega_{s(d)}^2 \cos^2p_x, \\
& \displaystyle {\tilde \gamma}_{5(6)} = 8V^4\frac{1}{N} \sum_p I_7(p) 
\omega_{s(d)}^2 \cos^2p_x, \\
& \displaystyle {\tilde \gamma}_7 = 32V^4\frac{1}{N} \sum_p I_6(p) 
\omega_s\omega_d \cos^2p_x, \\
& \displaystyle {\tilde \gamma}_8 = 16V^4\frac{1}{N} \sum_p I_7(p) 
\omega_s\omega_d \cos^2p_x, \\
& \displaystyle {\tilde \gamma}_{9(10)} = 32V^4\frac{1}{N} \sum_p I_6(p) 
\omega_{s(d)}^2 \cos p_x \cos p_y, \\
& \displaystyle {\tilde \gamma}_{11(12)} = 16V^4\frac{1}{N} \sum_p I_7(p) 
\omega_{s(d)}^2 \cos p_x \cos p_y, \\
&  \displaystyle  {\tilde \alpha}_m = U\Big(1+\frac{U}{N}\sum_p I_8(p)\Big), \\
&  \displaystyle  {\tilde \beta}_m = \frac{U^4}{2N}\sum_p I_9(p), \\
&  \displaystyle  {\tilde K}_m = \frac{U^2}{N}\sum_p I_9(p) 
\Big(\frac{\partial \xi_p}{\partial p_x}\Big)^2,  \\
&  \displaystyle  \{{\tilde \gamma}_{ms}, \ {\tilde \gamma}_{md}\} 
 = -4V^2U^2 \frac{1}{N} \sum_p [2I_{10}(p)+I_7(p)]
\{ \omega_s^2, \omega_d^2\}, \\
&  \displaystyle  \{ 
 {\tilde \gamma}_{mp1}, \ {\tilde \gamma}_{mp2}\}                
 = -4V^2U^2 \frac{1}{N} \sum_p [2I_{11}(p)+I_7(p)]    
\{\cos^2p_x, \cos p_x\cos p_y\}, \\
 &  \displaystyle  \{{\tilde \gamma}_{spm}, \ {\tilde \gamma}_{dpm}\} 
 = 8V^2U \frac{1}{N} \sum_p I_{12}(p) \cos p_x 
\{\omega_s,\omega_d\},  
\end{array}\end{equation}
where the functions appearing in the integrands are defined as  
\begin{equation}\begin{array}{rl}
I_4(p)  = & \displaystyle 
T\sum_{\epsilon_n}G_0(p,-i\varepsilon_n)G_0(p+Q,i\varepsilon_n),  \\
%
I_5(p) = & \displaystyle 
T\sum_{\epsilon_n}G_0^2(p,i\varepsilon_n)G_0^2(p+Q,-i\varepsilon_n), \\
%
I_6(p) = & \displaystyle T\sum_{\epsilon_n}
G_0^2(p,i\varepsilon_n)G_0(p,-i\varepsilon_n)G_0(p+Q,-i\varepsilon_n), \\
%
I_7(p) = & \displaystyle 
T\sum_{\epsilon_n}G_0(p,i\varepsilon_n)G_0(p,-i\varepsilon_n)
G_0(p+Q,i\varepsilon_n)G_0(p+Q,-i\varepsilon_n), \\
%
I_8(p) = & \displaystyle 
T\sum_{\epsilon_n}G_0(p,i\varepsilon_n)G_0(p+Q,i\varepsilon_n), \\
%
I_9(p) = & \displaystyle 
T\sum_{\epsilon_n}G_0^2(p,i\varepsilon_n)G_0^2(p+Q,i\varepsilon_n), \\
%
I_{10}(p) = & \displaystyle T\sum_{\epsilon_n}
G_0^2(p,i\varepsilon_n)G_0(p,-i\varepsilon_n)G_0(p+Q,i\varepsilon_n), \\
%
I_{11}(p) = & \displaystyle T\sum_{\epsilon_n}
G_0^2(p,i\varepsilon_n)G_0(p+Q,i\varepsilon_n)G_0(p+Q,-i\varepsilon_n),  \\
%
I_{12}(p) = & \displaystyle T\sum_{\epsilon_n}
G_0(p,i\varepsilon_n)G_0(p,-i\varepsilon_n)G_0(p+Q,i\varepsilon_n). \\

\end{array}\end{equation}

\end{document}